\documentclass[aps,twocolumn,showpacs,a4paper]{revtex4}
\usepackage{graphics}

\newcommand\bin[2]{{#1 \choose #2}}

\newcommand\ra{x}
\newcommand\rb{y}

\newcommand\eref[1]{(\ref{#1})}

\begin{document}

\title
{
  RNA-like polymer model: exact calculation on the Bethe lattice
}
\author{M. Pretti}
\affiliation
{
  Consiglio Nazionale delle Ricerche - Istituto Nazionale per la Fisica della Materia (CNR-INFM) \\
  Dipartimento di Fisica, Politecnico di Torino,
  Corso Duca degli Abruzzi 24, I-10129 Torino, Italy
}
\date{\today}
\begin{abstract}
We consider a lattice polymer model (random walk), in which the
walk is allowed to visit lattice bonds at most twice. Such a model
might have some relevance to describe statistical properties of
RNA molecules. In order to mimic base pairing, we assign an
attractive energy term to each doubly-visited bond, and a further
contribution to each pair of consecutive doubly-visited bonds. The
latter term is expected to mimic the stacking effect, whereas no
effect of sequence, that is, of chemical specificity, is taken
into account. The phase diagram is worked out exactly on a Bethe
lattice, in a grand-canonical formulation. In the single molecule
limit, the system undergoes two different phase transitions, upon
decreasing temperature: a $\Theta$-like collapse from a swollen
``coil'' state to a ``molten'' state, with a low fraction of
doubly-visited bonds, and subsequently to a ``paired'' state, with
empty or doubly-visited bonds only. The stacking effect drives the
latter transition from second to first order.
\end{abstract}

\pacs
{
05.70.Fh,  % Phase transitions: general studies
64.60.Cn,  % Order-disorder transformations; statistical mechanics of model systems
61.25.Hq,  % Macromolecular and polymer solutions; polymer melts; swelling
87.15.Aa,  % Biological and medical physics: Theory and modeling; computer simulation
87.14.Gg   % Biological and medical physics: DNA, RNA
}

\maketitle

\section{Introduction}

Lattice self-avoiding walks, i.e., random walks that are forbidden
to visit lattice sites more than once, have long been employed for
modeling linear polymers in a good solvent~\cite{Vanderzande1998}.
A short range interaction between nonconsecutive monomers has been
also considered, in order to represent either Van der Waals
attractive forces between monomers or the effective result of
solvophobic interactions. Such interactions cause the well-known
$\Theta$~transition from a swollen coil at high temperature to a
compact globule at low temperature~\cite{DeGennes1988}. The Bethe
approximation~\cite{Bethe1935,Kikuchi1951,Gujrati1995}, i.e., the
exact solution on the Bethe lattice, has been shown to reproduce
with reasonable accuracy, and with negligible computational
effort, the phase behavior of such basic model ($\Theta$~model)
and also of slightly more complicated polymer
models~\cite{Gujrati1995pol,Gujrati1998i,Gujrati1998ii,LiseMaritanPelizzola1998,BuzanoPretti2002}.

Several variations of the $\Theta$~model have been proposed, in
order to describe different physical phenomena. In particular, it
is possible to relax the self-avoidance constraint, allowing the
walk to visit lattice bonds at most twice. If the polymer chain is
assigned an orientation, and lattice bonds are allowed to be
doubly-visited only by opposite chain segments, the model is
denoted as 2-tolerant
trail~\cite{BaiesiOrlandiniStella2003,LeoniVanderzande2003}. This
model may be useful to investigate configurational statistics of
RNA molecules, whose importance in molecular biology is being more
and more
recognized~\cite{TinocoBustamante1999,Turner2002,DoudnaCech2002,MooreSteitz2002}.
Similar to DNA, a RNA molecule is a long polymer chain composed of
four different monomers (bases), adenine, cytosine, guanine, and
uracil, which are pairwise complementary (i.e., adenine-uracil and
cytosine-guanine pairings are energetically favored by formation
of hydrogen bonds~\cite{SponerLeszczynskiHobza2001}). At a
coarse-grained level, one can neglect the differences among bases,
and assign an attractive (contact) energy for each base pairing,
that is, for each doubly-visited bond.

Quite recently, Baiesi, Orlandini, and
Stella~\cite{BaiesiOrlandiniStella2003} have investigated the
previously described model, performing accurate Monte Carlo
simulations on the face-centered cubic (fcc) lattice, fully taking
into account the excluded volume effect, and showing the existence
of a continuous phase transition (similar to the
$\Theta$~collapse) from a high temperature state in which the RNA
is almost completely unpaired to a low temperature state with a
significant fraction of paired bases (the so-called molten phase).

In this work, we first verify that the Bethe approximation, which
is able to take into account excluded volume at a local
level~\cite{LiseMaritanPelizzola1998}, predicts a $\Theta$-like
transition as well. Moreover, we consider an extended model with a
more general energy function: We assign a specific energy
contribution to consecutive paired bases, without intermediate
branching, in order to mimic the so-called stacking
effect~\cite{TazawaKoikeInoue1980,ChenChenWentworth1990,SponerLeszczynskiHobza2001}.
Indeed, the stacking effect, mostly related to
hydrophobicity~\cite{TazawaKoikeInoue1980}, is claimed to be
energetically more relevant than base
pairing~\cite{SponerLeszczynskiHobza2001}, and however has great
importance for algorithms attempting to predict the secondary
structure of given RNA sequences~\cite{DimaHyeonThirumalai2005}.
In the statistical physics literature, models with ordinary
pairing energy
only~\cite{BundschuhHwa1999,BundschuhHwa2002,BaiesiOrlandiniStella2003,LeoniVanderzande2003,PagnaniParisiRicci2000,MarinariPagnaniRicci2002},
or with stacking energy
only~\cite{MukhopadhyayEmberlyTangWingreen2003}, or
both~\cite{Mueller2003} have been considered, but the relative
importance of the stacking effect with respect to base pairing has
been scarcely investigated~\cite{BurghardtHartmann2005}. On the
contrary, we specifically address the issue of stacking, assigning
different relevance to one of the two interactions, by means of an
adjustable parameter. We observe that our model predicts, in the
low temperature region, a fully base-paired phase. Such phase,
which does not at all correspond to a unique secondary structure,
might describe --with some cautions-- an average ``native'' state.
The phase transition to the molten phase (``denaturation'') turns
out to be continuous, for the ordinary pairing energy model, but,
upon adding even a small stacking energy, it turns out to change
into first order.

The paper is organized as follows. In
Sec.~\ref{sec:the_model_and_}, we introduce the model in some more
detail and give an overview of the Bethe lattice calculation. In
Sec.~\ref{sec:the_phase_} we work out the phase behavior of the
model, with particular attention to the single-molecule limit, and
in Sec.~\ref{sec:conclusions} we discuss the results, adding some
concluding remarks. Appendices~\ref{app:bethe} and~\ref{app:rec}
are devoted respectively to a derivation of the equilibrium free
energy and of the recursion equations for the Bethe lattice, while
in Appendix~\ref{app:theta} we report the analytical calculation
of the $\Theta$-like transition temperature.

\section{The model and the Bethe lattice calculation}
\label{sec:the_model_and_}

As previously mentioned, our polymer model is a self-avoiding
walk, which is exceptionally allowed to visit each lattice bond
(at most) twice, but is not allowed to self-intersect. We can
imagine roughly the following picture. Each segment of the walk
represents a monomer (base), whereas empty lattice bonds represent
the solvent. Doubly-visited bonds (i.e., bonds occupied by two
segments) represent paired bases, yielding an attractive
energy~$-(\beta-\gamma)$, with $\beta>\gamma>0$. Moreover, every
pair of consecutive doubly-visited bonds yields an additional
attractive contribution $-\gamma$, providing a rough description
of the stacking effect. It turns out that $\beta$ is the maximum
pairing energy, obtained by consecutive base-pairing, and will be
taken as the energy unit. According to the grand-canonical
formulation, a chemical potential~$\mu$ is associated to each
monomer, while the solvent chemical potential is conventionally
assumed to be zero.

Let us spend a few words to notify that the coarse-grained model
introduced above includes some degree of inconsistency. In
particular, to be more precise, segments of the walk ought to
represent stretches of the order of the persistence length of the
polymer chain. Unfortunately, the persistence length turns out to
be much different for single- or double-stranded RNA stretches
(being much larger in the latter case), but we nonetheless
describe both cases within a single lattice bond. We expect that
such inconsistency should not alter the qualitative phase behavior
of the model, since this is what happens for similar models of
DNA, as it has been also noted in
Ref.~\onlinecite{BaiesiOrlandiniStella2003}.

A configuration of the system can be defined by specifying the
number of segments on each lattice bond. Therefore, we define a
configuration variable $n_i=0,1,2$ (occupation number) for the
$i$-th lattice bond. Of course, such configuration variables are
not independent, but have to satisfy some constraints. In
particular, on each set of bonds of a given site, we impose the
following conditions: (i) the total number of segments must be
even; (ii) there cannot be more than 2 unpaired segments; (iii) if
only 2 segments are present, they must be unpaired.
Tab.~\ref{tab:energie} exemplifies the constraints in more detail,
for the simple case with coordination number equal to~$4$, but
generalizing to any coordination number is straightforward.

Constraint (i) is a simple connectivity constraint, stating that
the chain does not terminate after a finite number of segments.
Constraints (ii) and (iii) state that, if 2 unpaired segments come
to a given site from different lattice bonds, they either pair
each other (so that at least another bond is doubly occupied) or
they are consecutive along the chain (all other bonds are empty).
More precisely, constraint (ii) implies that unpaired chain
stretches behave like self-avoiding walks, which cannot visit a
lattice site more than once, unless they get paired. Constraint
(iii) deserves some more discussion. A configuration with only two
paired segments could represent a ``terminal loop'', in which the
chain bends onto itself, to form a hairpin. In principle, such
``zero-length loops'' should be allowed by a basic 2-tolerant
polymer model. Therefore, constraints (iii), which forbids them,
can be considered either as a further detail, which defines a
slightly different model, or as an approximation to the original
one. Such approximation, which is conceptually independent of the
subsequent approximate (Bethe) statistical treatment, has been
firstly taken for technical reasons, in order to simplify the
analytical calculations. We shall shortly discuss this technical
issue in the following. By now, we only observe that there is
actually a physical argument, which suggests that the modified
model might be even a bit closer to the real system. In fact, for
energetic reasons, terminal loops must have a minimum length of
four bases, and experiments show that, in real RNA, typical
hairpin loops are just of this kind
(tetraloops)~\cite{TinocoBustamante1999}. Therefore, a hairpin
loop should have a finite, though small, entropy, which cannot be
taken into account by a zero-length loop in the coarse-grained
model.

\begin{table}
  \setlength{\unitlength}{0.6mm}
  \caption
  {
    Configurations of a set of bonds of a given lattice site (left column);
    occupation numbers $n_i$ for each bond $i=0,\dots,k$ (mid columns);
    total number of segments $N_{n_0, \dots, n_k} \equiv \sum_{i=0}^k n_i$
    and corresponding energy term $H_{n_0, \dots, n_k}$ (right columns).
    Notice that: graphical representations are limited to four
    bonds; only configurations with even number of segments are
    reported, because odd numbers are forbidden
    (the corresponding energy terms are~$\infty$);
    energy terms are invariant under bond permutations.
  }
  \begin{ruledtabular}
  \begin{tabular}{c|ccccccc|rr}
    conf.
    & $n_0$ & $n_1$ & $n_2$ & $n_3$ & $n_4$ & $\dots$ & $n_k$ & $N_{n_0, \dots, n_k}$ & $H_{n_0, \dots, n_k}$ \cr
    \hline
    \begin{picture}(16,14)(-8,-2)
      %\put(-8,-8){\framebox(16,16)}
      \thinlines
      \put( 2, 0){\line( 1, 0){5}}
      \put(-2, 0){\line(-1, 0){5}}
      \put( 0, 2){\line( 0, 1){5}}
      \put( 0,-2){\line( 0,-1){5}}
    \end{picture}
    & $0$   & $0$   & $0$   & $0$   & $0$   & $\dots$ & $0$   & $0$ & $0$ \cr
    \begin{picture}(16,16)(-8,-2)
      %\put(-8,-8){\framebox(16,16)}
      \thinlines
      \put(-2, 0){\line(-1, 0){5}}
      \put( 0, 2){\line( 0, 1){5}}
      \put( 0,-2){\line( 0,-1){5}}
      \thicklines
      \put(0, 0.5){\line( 1, 0){7}}
      \put(0, 0.7){\line( 1, 0){7}}
      \put(0,-0.5){\line( 1, 0){7}}
      \put(0,-0.7){\line( 1, 0){7}}
    \end{picture}
    & $2$   & $0$   & $0$   & $0$   & $0$   & $\dots$ & $0$   & $2$ & $\infty$ \cr
    \begin{picture}(16,16)(-8,-2)
      %\put(-8,-8){\framebox(16,16)}
      \thinlines
      \put(-2, 0){\line(-1, 0){5}}
      \put( 0,-2){\line( 0,-1){5}}
      \thicklines
      \put(-0.1, 0.1){\line( 1, 0){7.1}}
      \put(-0.1,-0.1){\line( 1, 0){7.1}}
      \put( 0.1,-0.1){\line( 0, 1){7.1}}
      \put(-0.1,-0.1){\line( 0, 1){7.1}}
    \end{picture}
    & $1$   & $1$   & $0$   & $0$   & $0$   & $\dots$ & $0$   & $2$ & $0$ \cr
    \begin{picture}(16,16)(-8,-2)
      %\put(-8,-8){\framebox(16,16)}
      \thinlines
      \put(-2, 0){\line(-1, 0){5}}
      \put( 0,-2){\line( 0,-1){5}}
      \thicklines
      \put( 0.5, 0.5){\line( 1, 0){6.5}}
      \put( 0.7, 0.7){\line( 1, 0){6.3}}
      \put(-0.5,-0.5){\line( 1, 0){7.5}}
      \put(-0.7,-0.7){\line( 1, 0){7.7}}
      \put( 0.5, 0.5){\line( 0, 1){6.5}}
      \put( 0.7, 0.7){\line( 0, 1){6.3}}
      \put(-0.5,-0.5){\line( 0, 1){7.5}}
      \put(-0.7,-0.7){\line( 0, 1){7.7}}
    \end{picture}
    & $2$   & $2$   & $0$   & $0$   & $0$   & $\dots$ & $0$   & $4$ & $-\gamma$ \cr
    \begin{picture}(16,16)(-8,-2)
      %\put(-8,-8){\framebox(16,16)}
      \thinlines
      \put( 0,-2){\line( 0,-1){5}}
      \thicklines
      \put( 0.5, 0.5){\line( 1, 0){6.5}}
      \put( 0.7, 0.7){\line( 1, 0){6.3}}
      \put(   0,-0.5){\line( 1, 0){7}}
      \put(   0,-0.7){\line( 1, 0){7}}
      \put( 0.5, 0.5){\line( 0, 1){6.5}}
      \put( 0.7, 0.7){\line( 0, 1){6.3}}
      \put(   0,-0.5){\line(-1, 0){7}}
      \put(   0,-0.7){\line(-1, 0){7}}
    \end{picture}
    & $2$   & $1$   & $1$   & $0$   & $0$   & $\dots$ & $0$   & $4$ & $0$ \cr
    \begin{picture}(16,16)(-8,-2)
      %\put(-8,-8){\framebox(16,16)}
      \thicklines
      \put(   0, 0.1){\line( 1, 0){7}}
      \put(   0,-0.1){\line( 1, 0){7}}
      \put( 0.1,   0){\line( 0, 1){7}}
      \put(-0.1,   0){\line( 0, 1){7}}
      \put(   0, 0.1){\line(-1, 0){7}}
      \put(   0,-0.1){\line(-1, 0){7}}
      \put( 0.1,   0){\line( 0,-1){7}}
      \put(-0.1,   0){\line( 0,-1){7}}
    \end{picture}
    & $1$   & $1$   & $1$   & $1$   & $0$   & $\dots$ & $0$   & $4$ & $\infty$ \cr
    \begin{picture}(16,16)(-8,-2)
      %\put(-8,-8){\framebox(16,16)}
      \thinlines
      \put( 0,-2){\line( 0,-1){5}}
      \thicklines
      \put( 0.5, 0.5){\line( 1, 0){6.5}}
      \put( 0.7, 0.7){\line( 1, 0){6.3}}
      \put(-0.5, 0.5){\line( 0, 1){6.5}}
      \put(-0.7, 0.7){\line( 0, 1){6.3}}
      \put(   0,-0.5){\line( 1, 0){7}}
      \put(   0,-0.7){\line( 1, 0){7}}
      \put( 0.5, 0.5){\line( 0, 1){6.5}}
      \put( 0.7, 0.7){\line( 0, 1){6.3}}
      \put(-0.5, 0.5){\line(-1, 0){6.5}}
      \put(-0.7, 0.7){\line(-1, 0){6.3}}
      \put(   0,-0.5){\line(-1, 0){7}}
      \put(   0,-0.7){\line(-1, 0){7}}
    \end{picture}
    & $2$   & $2$   & $2$   & $0$   & $0$   & $\dots$ & $0$   & $6$ & $0$ \cr
    \begin{picture}(16,16)(-8,-2)
      %\put(-8,-2){\framebox(16,16)}
      \thicklines
      \put( 0.5, 0.5){\line( 1, 0){6.5}}
      \put( 0.7, 0.7){\line( 1, 0){6.3}}
      \put( 0.5, 0.5){\line( 0, 1){6.5}}
      \put( 0.7, 0.7){\line( 0, 1){6.3}}
      \put(-0.5, 0.5){\line( 0, 1){6.5}}
      \put(-0.7, 0.7){\line( 0, 1){6.3}}
      \put(-0.5, 0.5){\line(-1, 0){6.5}}
      \put(-0.7, 0.7){\line(-1, 0){6.3}}
      \put( 0.5,-0.5){\line( 0,-1){6.5}}
      \put( 0.7,-0.7){\line( 0,-1){6.3}}
      \put( 0.5,-0.5){\line( 1, 0){6.5}}
      \put( 0.7,-0.7){\line( 1, 0){6.3}}
    \end{picture}
    & $2$   & $2$   & $1$   & $1$   & $0$   & $\dots$ & $0$   & $6$ & $0$ \cr
    &&&&&&&&& \cr
    \vdots
    & $\vdots$ & $\vdots$ & $\vdots$ & $\vdots$ & $\vdots$ & $\ddots$ & $\vdots$ & $\vdots$ & $\vdots$ \cr
    \begin{picture}(16,16)(-8,-2)
      %\put(-8,-8){\framebox(16,16)}
      \thicklines
      \put( 0.5, 0.5){\line( 1, 0){6.5}}
      \put( 0.7, 0.7){\line( 1, 0){6.3}}
      \put( 0.5, 0.5){\line( 0, 1){6.5}}
      \put( 0.7, 0.7){\line( 0, 1){6.3}}
      \put(-0.5, 0.5){\line( 0, 1){6.5}}
      \put(-0.7, 0.7){\line( 0, 1){6.3}}
      \put(-0.5, 0.5){\line(-1, 0){6.5}}
      \put(-0.7, 0.7){\line(-1, 0){6.3}}
      \put(-0.5,-0.5){\line(-1, 0){6.5}}
      \put(-0.7,-0.7){\line(-1, 0){6.3}}
      \put(-0.5,-0.5){\line( 0,-1){6.5}}
      \put(-0.7,-0.7){\line( 0,-1){6.3}}
      \put( 0.5,-0.5){\line( 0,-1){6.5}}
      \put( 0.7,-0.7){\line( 0,-1){6.3}}
      \put( 0.5,-0.5){\line( 1, 0){6.5}}
      \put( 0.7,-0.7){\line( 1, 0){6.3}}
    \end{picture}
    & $2$   & $2$   & $2$   & $2$   & $2$   & $\dots$ & $2$   & $2(k+1)$ & $0$ \cr
    &&&&&&&&& \cr
  \end{tabular}
  \end{ruledtabular}
  \label{tab:energie}
\end{table}

Assuming a coordination number~$k+1$, the Hamiltonian can be
formally written as
\begin{equation}
  \mathcal{H} =
  \sum_{\{ i_0, \dots, i_k \}}
  H_{n_{\scriptstyle i_0}, \dots, n_{\scriptstyle i_k}}
  + \sum_{i} h_{n_{\scriptstyle i}}
  ,
  \label{eq:hamiltonian}
\end{equation}
where the former sum runs over all sets of bonds $\{ i_0, \dots,
i_k \}$ of all lattice sites, and the latter over all bonds $i$.
Single-bond energy terms $h_n$ take into account pairing energies
and chemical potential contributions, and can be defined as
follows
\begin{eqnarray}
  h_0 & = & 0
  , \label{eq:h0} \\
  h_1 & = & - \mu
  , \label{eq:h1} \\
  h_2 & = & - (\beta - \gamma) - 2\mu
  . \label{eq:h2}
\end{eqnarray}
Many-bond terms $H_{n_0,\dots,n_k}$ take into account the
constraints, assigning infinite energy penalties to forbidden
configurations, and the stacking energy contributions. A
definition of these terms would be quite cumbersome, from an
analytical point of view, so that we can assume they are defined
by a table like Tab.~\ref{tab:energie}.

Let us briefly return to discuss constraint (iii), which, as
previously mentioned, disallows zero-length hairpin loops, i.e.,
configurations with only two paired segments on the set of bonds
of a given site. It turns out that, if we allowed such a
configuration, local constraints of the form $H_{n_0,\dots,n_k}$
would also allow, for instance, configurations with two paired
segments disconnected from everything else, or even paired
stretches of any length. In this way, we would not study an
infinitely long polymer, but a mixture of polymers of different
lengths, and this would be a completely different system. In order
to avoid constraint (iii), we would have to exclude the
aforementioned undesired configurations, and we would need a more
complicated treatment. This will be the subject of a future work.

Let us now introduce the Bethe approximation. Basically, it
consists in replacing the single walk on the regular lattice by a
gas of walks, with the same self-avoidance constraints and
interactions, on a Bethe lattice, having the same coordination
number as the regular lattice one. In older literature, a Bethe
lattice was simply understood to be the inner region of an
infinite Cayley tree~\cite{Domb1960}. Nevertheless, it has been
subsequently shown that the thermodynamic behavior of a Cayley
tree is strongly affected by the presence of a boundary, and the
exact solution for this system does not agree with the Bethe
approximation~\cite{Eggarter1974}. More recently, it has been
recognized that the Bethe lattice has to be a homogeneous,
boundary-less structure (so that the thermodynamic properties of
the system can be worked out by suitable self-consistence
equations~\cite{Gujrati1995}), which also allows for the presence
of macroscopically large loops~\cite{GujratiBowman1999}. The Bethe
lattice is thus better defined as an ensemble of random graphs
with fixed coordination number~\cite{MezardParisi2001}, which are
locally treelike (in the sense that loop length is
$\mathcal{O}(\ln \mathcal{N})$, where $\mathcal{N}$ is the number
of nodes), and whose thermodynamic behavior is governed by the
variational Bethe free energy~\cite{Kikuchi1951}. The free energy
minima can still be determined by solving a suitable recursion
relation for the so-called partial partition
functions~\cite{Gujrati1995,Pretti2003}. We shall address this
issue in Appendix~\ref{app:bethe}. Hereafter, we just give an
intuitive derivation, based on the treelike nature of the system.
\begin{figure}
  \resizebox{80mm}{!}{\includegraphics*[10mm,205mm][110mm,265mm]{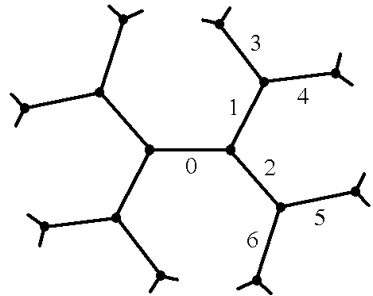}}
  \caption
  {
    Sketch of a Bethe lattice with $k=2$.
  }
  \label{fig:bethelattice}
\end{figure}
Let us consider for instance the Bethe lattice depicted in
Fig.~\ref{fig:bethelattice}, and the right part of the system,
starting with the bond denoted by~$0$. Since loops connecting the
two parts are (with high probability) infinitely long in the
thermodynamic limit, we can imagine that the two parts are
actually disconnected branches and that we can thus define a
partial Hamiltonian, obtained by Eq.~(\ref{eq:hamiltonian})
restricting the sum to bond variables in one branch. The
corresponding partial partition function~$W_{n_0}$ can be computed
by summing the Boltzmann weights of the partial Hamiltonian over
the configurations of the branch except the $0$~bond. Actually, it
is convenient to work with a normalized partial partition function
$w_{n_0} \propto W_{n_0}$, such that
\begin{equation}
  \sum_{n_0=0}^2 w_{n_0} = 1
  .
  \label{eq:norm}
\end{equation}
The normalized partial partition function $w_{n_0}$ represents, as
a function of $n_0$, the probability distribution of the
configuration variable ``in the absence of the other branch''. Let
us now observe that, in the thermodynamic limit, and in the
hypothesis of a homogeneous system, the subbranches attached to
the $0$ bond should be equivalent to main one, so that one can
write the recursion equation
\begin{equation}
  w_{n_0} = q^{-1} e^{-h_{n_0}}
  \sum_{n_1=0}^{2} \dots \sum_{n_k=0}^{2}
  e^{-H_{n_0,n_1,\dots,n_k}}
  \prod_{i=1}^{k} w_{n_i}
  .
  \label{eq:rec}
\end{equation}
The sum runs over configuration variables of bonds attached to the
$0$ bond ($n_1$ and $n_2$, in our example), the energy terms
$h_{\scriptstyle n_0}$ and $H_{n_0,n_1,\dots,n_k}$ are assumed to
be normalized to temperature, and $q$~is a normalization constant,
imposed by Eq.~(\ref{eq:norm}). A more explicit form of the
recursion equation is given in Appendix~\ref{app:rec}, where the
specific energy terms $H_{n_0,n_1,\dots,n_k}$ of our model are
taken into account. The recursion equation can be solved
numerically by a simple iterative algorithm. All equilibrium
properties of the system can be derived from the knowledge of the
partial partition function.

First of all, we can compute the probability distribution~$p_n$ of
a bond configuration variable~$n$, by considering the operation of
attaching $2$~branches to the given bond. We obtain
\begin{equation}
  p_n = z^{-1} e^{h_n} w_n^2
  ,
  \label{eq:pjoint}
\end{equation}
where
\begin{equation}
  z = \sum_{n=0}^2 e^{h_n} w_n^2
  \label{eq:zjoint}
\end{equation}
provides normalization. The average number of segments per bond,
which we shall briefly refer to as {\em density} in the following,
can be evaluated as
\begin{equation}
  \rho = \sum_{n=0}^2 n p_n  = p_1 + 2 p_2
  .
\end{equation}
The density~$\rho$ is the main order parameter for our system. As
a secondary order parameter, we evaluate the fraction of paired
segments
\begin{equation}
  \phi = 2 p_2 / \rho
  .
\end{equation}
The grand-canonical free energy (grand-potential) per
bond~$\omega$ can be determined as
\begin{equation}
  \omega = - \frac{2 \ln q - (k-1) \ln z}{k+1}
  ,
  \label{eq:enlib}
\end{equation}
where $q$~is the normalization constant of the recursion
equation~\eref{eq:rec} and $z$~is given by Eq.~\eref{eq:zjoint}.
The derivation of this expression requires some manipulations and
is reported in Appendix~\ref{app:bethe}. From the knowledge of the
grand-potential one can derive all other thermodynamic properties,
and determine thermodynamic stability for each phase.

\section{The phase diagram and the single-molecule limit}
\label{sec:the_phase_}

In the framework of a grand-canonical formulation, the phase
diagram can be described as a function of temperature and chemical
potential. For a polymer, the latter controls the average chain
length. For example, in the simple $\Theta$~model, there exists a
phase transition line $\mu = \varphi(T)$ at which (for increasing
$\mu$ values) the average length either diverges continuously (for
temperatures higher than some temperature $\Theta$) or jumps
discontinuously to infinity (for temperatures lower than
$\Theta$)~\cite{Vanderzande1998}. The transition line is
identified as the thermodynamic limit of a single chain, so that
we denote it as the ``single-molecule'' limit. Alternatively, the
system can be described in terms of a segment density~$\rho$, and
one obtains $\rho=0$ for $\mu<\varphi(T)$ and $\rho>0$ for
$\mu>\varphi(T)$. The transition is second order for $T>\Theta$
and first order for $T<\Theta$. In the limit $\mu\to\varphi(T)^+$,
the properties of the dense phase approach those of a single
chain, and, in particular, the segment density~$\rho$ is a measure
of the chain compactness. Therefore, the tricritical point
$(\Theta,\varphi(\Theta))$, known as $\Theta$~point, represents a
coil-globule collapse.

\begin{figure}
  \resizebox{80mm}{!}{\includegraphics*[30mm,70mm][180mm,245mm]{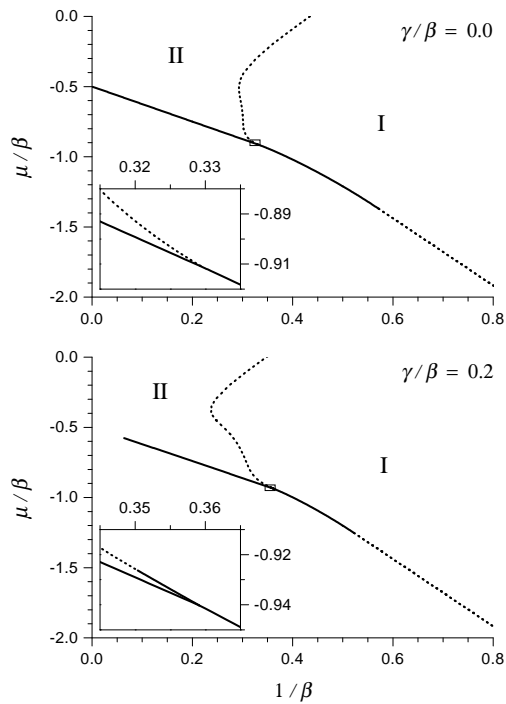}}
  \caption
  {
    Chemical potential-temperature ($\mu/\beta$ vs $1/\beta$) phase diagram
    for the zero stacking case ($\gamma = 0$, upper graph)
    and a nonzero stacking case ($\gamma/\beta = 0.2$, lower graph).
    Solid lines denote first order transitions;
    dashed lines denote second order transitions.
    The ordinary dense phase is denoted by I,
    whereas the fully paired dense phase is denoted by II.
    The zero density phase is left blank.
    The insets display the regions enclosed in the small rectangles.
  }
  \label{fig:tmu}
\end{figure}
We present grand-canonical phase diagrams of our Bethe lattice
model for the case of zero stacking effect ($\gamma = 0$) and for
a case of nonzero stacking effect ($\gamma/\beta = 0.2$), which
show qualitatively different behaviors. Let us recall that
$\gamma/\beta$ quantifies the ratio between the neat stacking
energy $\gamma$ and the total effect of pairing and stacking (the
simple pairing energy is $\beta-\gamma$). We shall shortly denote
$\gamma/\beta$ as stacking ratio in the following. We set $k=11$
(coordination number $=12$), expecting to approximate the fcc
lattice. Let us consider the zero stacking case first. The phase
diagram is displayed in Fig.~\ref{fig:tmu} (upper graph), where
the temperature variable is~$1/\beta$ and the chemical potential
variable is~$\mu/\beta$. We find three different phases: a zero
density phase (O), an ordinary dense phase (I), and a fully paired
dense phase (II). The zero density phase is characterized by
$\rho=0$. Since only a vanishing fraction of bonds is occupied in
this phase, also the grand-potential per bond $\omega$ vanishes,
and, for the same reason, the fraction of paired segments $\phi$
is undefined. The I phase is characterized by $0<\rho<2$ and
$0<\phi<1$, i.e., it is a dense phase which possesses a finite
fraction of paired segments. We can roughly compare it to the
dense phase of an ordinary $\Theta$~model. Finally, the II phase
is characterized by $0<\rho<2$ and $\phi=1$, i.e., it is a dense
phase in which every segment is paired.

The transition line between the O and I phases turns out to be
partially first and partially second order. The two regimes are
separated by a tricritical point. As explained in
Appendix~\ref{app:theta}, it is possible to determine analytically
the equation of the second order line,
\begin{equation}
  \mu = - \ln k
  ,
  \label{eq:crit}
\end{equation}
and the location of the tricritical point
\begin{equation}
  \beta = \ln \frac{k^2}{k + (k-1) \, e^{-\gamma}}
  .
  \label{eq:tric}
\end{equation}
It has been pointed out that, for this kind of (2-tolerant)
polymer models, the tricritical point exhibits peculiar values of
the critical exponents, that are different from those of the
ordinary self-avoiding walk with attractive interaction, and
suggest a linear-to-branched polymer
transition~\cite{OrlandiniSenoStellaTesi1992}. Evidences of such a
behavior will be observed also in our model. Nevertheless, since
this point still corresponds to a continuous collapse and, since
exponent differences cannot be detected at a Bethe approximation
level, we shall all the same speak of a $\Theta$~point in the
following.

In the dense region ($\rho>0$) a second order transition line
separates the I and II phases. This line joins to the transition
line with the O phase at a critical end-point. The latter
corresponds to another continuous conformational transition for
the single molecule. The behavior is different, in the presence of
the stacking effect, as shown in Fig.~\ref{fig:tmu} (lower graph).
The same three phases O, I, and II discussed above are present,
and also the high temperature region of the phase diagram is
qualitatively similar, although we can observe a lower
$\Theta$~temperature, in agreement with Eq.~\eref{eq:tric}. On the
contrary, the I-II transition line turns out to be partially
second and partially first order, giving rise to a tricritical
point in the dense region. In this way, the critical end-point
disappears, and is replaced by a triple point, which corresponds
to a discontinuous transition in the single-molecule limit.

\begin{figure}
  \resizebox{80mm}{!}{\includegraphics*[30mm,155mm][180mm,245mm]{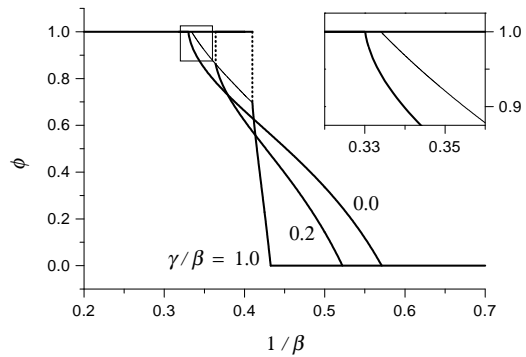}}
  \caption
  {
    Fraction of paired segments as a function of temperature ($\phi$ vs $1/\beta$)
    in the single-molecule limit,
    for different values of the stacking ratio $\gamma/\beta$.
    Dashed lines denote discontinuities;
    a thin solid line connects transition values.
    The inset displays the region enclosed in the small rectangle.
  }
  \label{fig:tphi}
\end{figure}
Let us now investigate this limit in more detail. First of all, we
consider the fraction of paired segments~$\phi$, computed for
$\mu$ tending to the transition line from above, as a function of
temperature. The results are reported in Fig.~\ref{fig:tphi}, for
three different values of the stacking ratio
$\gamma/\beta=0,0.2,1$. For all cases, we can see that $\phi$ is
rigorously zero above the $\Theta$~temperature. In this regime,
which we can denote as {\em coil} state, the polymer behaves like
an ordinary self-avoiding walk without self-interaction. Upon
decreasing temperature below the $\Theta$~point, the fraction of
paired segments begins to increase, revealing formation of
contacts. We can identify this regime as the {\em molten} state.
As previously mentioned, the $\Theta$ temperature decreases, upon
increasing the stacking energy. Upon further decreasing
temperature, $\phi$ reaches the saturation value $\phi=1$. In this
regime, which we simply denote as {\em paired} state since all
segments are paired, we can imagine our system as a branched
double chain. In this sense, we can identify this phase as a
``native'' RNA-like state, although it does not at all correspond
to a single configuration, as it will become clearer later. As
previously mentioned, the molten-paired transition is continuous
in the zero stacking case, but becomes first order in the nonzero
stacking cases. More precisely, we observe that the stacking
energy needed to change the order of the transition is very small
but finite, as suggested in Fig.~\ref{fig:tphi} by the thin line
connecting the transition values of $\phi$.

\begin{figure}
  \resizebox{80mm}{!}{\includegraphics*[30mm,155mm][180mm,245mm]{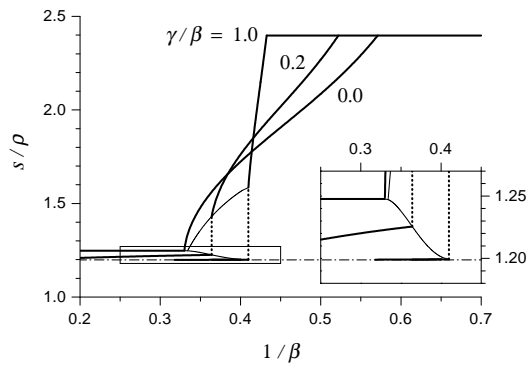}}
  \caption
  {
    Entropy per segment as a function of temperature
    ($s/\rho$ vs $1/\beta$)
    in the single-molecule limit,
    for different values of the stacking ratio $\gamma/\beta$.
    Dashed lines denote discontinuities;
    thin solid lines connect transition values.
    The inset displays the region enclosed in the small rectangle.
    A dash-dotted line indicates the entropy value $\ln k/2$.
  }
  \label{fig:tsigma}
\end{figure}
We also investigate the temperature dependence of the entropy per
segment, which can be computed as follows. The grand-potential per
bond can be written as
\begin{equation}
  \omega = f - \mu \rho
  ,
\end{equation}
where $f$ is the Helmholtz free energy per bond. As previously
mentioned, $\omega$ vanishes at the O~phase boundary, so that in
this case $\mu$ coincides with the Helmholtz free energy per
segment
\begin{equation}
  \mu = f/\rho
  .
\end{equation}
Remembering that our energies are normalized to temperature, the
equation of the O~phase boundary can be written as
\begin{equation}
  \mu/\beta = \varphi (1/\beta)
  ,
  \label{eq:phi}
\end{equation}
where the function~$\varphi$ is known numerically with high
precision. The entropy per segment (in natural units) can thus be
easily determined as
\begin{equation}
  s/\rho = -\varphi'(1/\beta)  ,
  \label{eq:phiprime}
\end{equation}
where $\varphi'$ denotes the first derivative of $\varphi$. We
report the results in Fig.~\ref{fig:tsigma}, for the usual values
of the stacking ratio $\gamma/\beta=0,0.2,1$. For all cases, in
the coil state, the entropy per segment is rigorously independent
of temperature and equal to $\ln k$, as it can be easily derived
by Eqs.~\eref{eq:phi}, \eref{eq:crit} and~\eref{eq:phiprime}. This
value characterizes an ordinary self-avoiding walk on the Bethe
lattice~\cite{LiseMaritanPelizzola1998}. In the molten state, the
entropy starts decreasing (as temperature decreases), more and
more rapidly, upon increasing the stacking effect. Finally, in the
paired state, the entropy is almost constant and its value turns
out to be slightly larger than $(\ln k)/2$, which would
characterize a self-avoiding double chain. The excess entropy with
respect to this value, which is due to branching, tends to zero as
temperature goes to zero, and decreases upon increasing the
stacking effect.

\begin{figure}
  \resizebox{80mm}{!}{\includegraphics*[30mm,155mm][180mm,245mm]{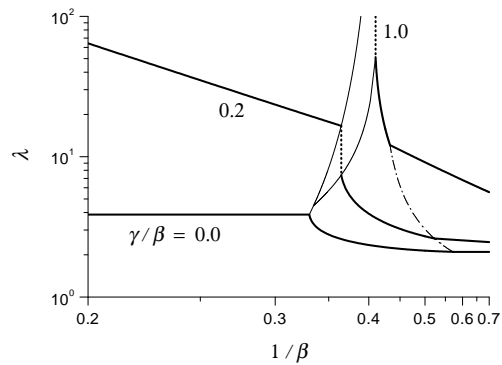}}
  \caption
  {
    Average stacking length as a function of temperature
    ($\lambda$ vs $1/\beta$)
    in the single-molecule limit,
    for different values of the stacking ratio $\gamma/\beta$.
    Dashed lines denote discontinuities;
    thin solid lines connect transition values;
    a dash-dotted line,
    determined by Eqs.~\eref{eq:stacklength} and~\eref{eq:pibeta},
    connects $\Theta$-point values.
  }
  \label{fig:tlambda}
\end{figure}
According to the results reported so far, the low temperature
phase might appear as almost completely quenched. The following
analysis of the average length of double chain stretches (which we
shall shortly refer to as stacking length in the following)
demonstrates that this is not the case. Let $\pi$ denote the
probability that, given a lattice bond occupied by two paired
segments, just one neighbor bond (in a given direction) is
occupied by two paired segments as well. We can call $\pi$ the
stacking probability. Due to the Markovian nature of the Bethe
lattice, the probability of having a double chain stretch of
length $l$ in the given direction is $\pi^{l-1}(1-\pi)$. The
average stacking length is therefore
\begin{equation}
  \lambda
  =
  \sum_{l=1}^{\infty} l \pi^{l-1} (1-\pi)
  =
  \frac{1}{1-\pi}
  .
  \label{eq:stacklength}
\end{equation}
Considering the recursion equation~\eref{eq:rec-2prev}, we can
derive the stacking probability as the ratio between the weight of
the stacked configuration $\bin{k}{1} e^{\gamma} w_2 w_0^{k-1}$
and the total weight of the configurations compatible with the two
paired segments (coinciding with the left-hand side, at a fixed
point of the recursion equations). Remembering also the
expression~\eref{eq:h2} for $h_2$, we easily obtain
\begin{equation}
  \pi = q^{-1} k \, e^{2\mu+\beta} w_0^{k-1}
  ,
\end{equation}
where of course $q$ and~$w_0$ are available from the numerical
solution. It is also useful to derive an explicit expression
for~$\pi$ at the second order O-I phase boundary, in order to
avoid taking limits numerically. Performing basically the same
calculation with Eq.~\eref{eq:rec-y}, taking the limit $x,y \to
0$, and making use of Eq.~\eref{eq:xvsy}, we obtain
\begin{equation}
  \pi = \frac{k}{k + (k-1) \, e^{-\gamma}}
  .
\end{equation}
Moreover, comparing this equation with Eq.~\eref{eq:tric}, we
obtain, at the $\Theta$~point, the following simple relation
\begin{equation}
  \pi = e^\beta/k
  .
  \label{eq:pibeta}
\end{equation}
The results are reported in Fig.~\ref{fig:tlambda}, again for
$\gamma/\beta=0,0.2,1$. The most interesting features appear in
the coil and paired states. In particular, we can observe that the
stacking length is constant with respect to temperature if
$\gamma=0$, i.e., in the absence of the stacking effect. On the
contrary, even a very small stacking energy makes the stacking
length increase upon decreasing temperature. Since the length
scale is logarithmic and the temperature scale is ``inverse'',
straight lines indicate that $\lambda$ is exponential in $\beta$
in these phases. As a consequence, in the presence of the stacking
effect, the stacking length diverges as temperature goes to zero,
so that the ground state of the model can be thought of as a
unique double chain (hairpin). Let us also notice that, in the
coil state, the stacking length does not vanish at any finite
temperature value, unlike the fraction of paired segments. These
results do not disagree, meaning that, if a (rare) contact is
formed, it has nevertheless a probability of not being isolated.

\begin{figure}
  \resizebox{80mm}{!}{\includegraphics*[30mm,155mm][180mm,250mm]{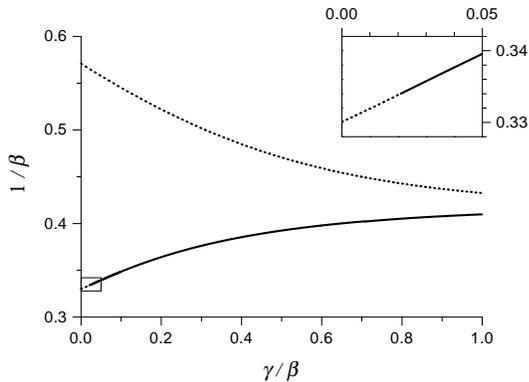}}
  \caption
  {
    Transition temperatures as a function of the stacking ratio
    ($1/\beta$ vs $\gamma/\beta$)
    in the single-molecule limit.
    Solid lines denote first order transitions;
    dashed lines denote second order transitions.
    The inset displays the region enclosed in the small rectangle.
  }
  \label{fig:stacking}
\end{figure}
We have already described the overall behavior of the phase
transitions in the single-molecule limit, as a function of the
stacking ratio. As this parameter increases, the
$\Theta$~temperature decreases, whereas the molten-paired
transition temperature increases, and the transition changes from
second to first order. For the sake of completeness, we report in
Fig.~\ref{fig:stacking} the transition temperatures as a function
of the stacking ratio. The $\Theta$~transition line is given
analytically by Eq.~\eref{eq:tric}, whereas the molten-paired
transition has been determined numerically. In the latter
transition, we recover the continuous regime at very low stacking
values and the discontinuous regime at higher stacking values.
Quantitatively, the boundary between the two regimes is found to
occur at $\gamma/\beta \approx 0.0210$.

\section{Discussion and conclusions}
\label{sec:conclusions}

In this paper, we have investigated a 2-tolerant polymer model on
the Bethe lattice, with both contact and stacking interactions.
The model is expected to mimic some qualitative features of
conformational transitions of RNA molecules. The most striking
results are the presence of a low temperature transition to a
fully-paired phase and the effect of stacking, which turns out to
drive such transition from second to first order. We have already
questioned, throughout the paper, whether these results have some
relevance to the denaturation transition of real RNA. The most
important warning concerns the fact that we completely neglect
chemical heterogeneities, which are indeed present in RNA. As a
consequence, we observe that the fully-paired state does not
correspond to a well defined secondary structure, but indeed to a
variety of structures. On the contrary, several investigations
proposed in the literature take into account randomly distributed
heterogeneous
sequences~\cite{BundschuhHwa1999,BundschuhHwa2002,PagnaniParisiRicci2000,MarinariPagnaniRicci2002,BurghardtHartmann2005}.
Even in this case, the low temperature glasslike phase does not
correspond to a fixed structure, but the general claim is that it
could describe average RNA properties. On this kind of models, the
only work we are aware of, which performs a systematic
investigation as a function of the strength of the stacking
interaction, is one by Burghardt and
Hartmann~\cite{BurghardtHartmann2005}. In the cited paper, the
authors do not find any evidence of a change in the order of the
(temperature-induced) denaturation transition. Nevertheless, in a
previous work, Zhou and Zhang~\cite{ZhouZhang2001} had observed
that an increasing stacking energy could change the order of a
force-induced denaturation. This result has been actually
criticized by M\"uller~\cite{Mueller2003}, who argued that the
apparent first order transition was rather to be interpreted as a
sharp cross-over. It is important to notice, however, that all the
cited investigations neglect the effect of excluded volume. On the
contrary, the Bethe lattice approximation is partially able to
account for excluded volume, by imposing local self-avoidance
constraints. This may be indeed a reason for our qualitatively
different results. In order to investigate this issue in more
detail, it would be interesting to extend the Bethe lattice
analysis to the case of a random heterogeneous sequence, making
use the recently proposed cavity method~\cite{MezardParisi2001},
along the lines traced by Montanari, M{\"u}ller, and Mezard, for
the self-avoiding
heteropolymer~\cite{MontanariMuellerMezard2004,MuellerMezardMontanari2004}.

Let us also compare our results with the Monte Carlo simulations
by Baiesi, Orlandini, and Stella~\cite{BaiesiOrlandiniStella2003}.
In the cited work, the authors investigate a 2-tolerant trail with
contact energy only, on the fcc lattice, fully taking into account
excluded volume. As the polymer is assigned an orientation (whence
the term ``trail''), only antiparallel contacts are allowed. Let
us notice, by the way, that in our treatment we have not
introduced orientation explicitly. Nevertheless, it is possible to
show that the latter, together with the constraint on antiparallel
contacts, are equivalent to a simple renormalization of the
partial partition functions, with no effect on observable
quantities. Of course, we cannot expect that any result concerning
critical exponents could be reproduced in the framework of our
mean-field-like approach. Nevertheless, it is noticeable that, for
$k=11$ (fcc lattice) and $\gamma=0$ (contact energy only), we
predict a $\Theta$~temperature ($\beta \approx 1.7513$) not so far
from the Monte Carlo result ($\beta \approx 1.9$). Apart from this
result, the fact itself that we observe a $\Theta$-like transition
may rise some interest. Indeed, an important result of
Ref.~\onlinecite{BaiesiOrlandiniStella2003} is that, if {\em
pseudoknots} (i.e., tertiary RNA structure) are forbidden, then
the $\Theta$-like transition is replaced by a smooth cross-over.
We wonder how the Bethe lattice approach can reproduce (at least
qualitatively) the situation {\em with} pseudoknots, since the
model is embedded on a treelike structure. Our tentative
explanation is based on the previously mentioned definition of the
Bethe lattice in terms of random graphs~\cite{MezardParisi2001},
rather than the more usual and quick definition (``the infinite
interior of the infinite Cayley tree''~\cite{Gujrati1995}). In the
former picture, pseudoknots could be realized via large loops
present in the graphs, although in the current treatment we do not
have any control on them. Let us finally notice that, conversely,
just the presence of only large loops in the Bethe lattice might
create some artifact in the description of RNA statistics,
especially with a constraints which disallows zero-length hairpin
turns, as discussed in Sec.~\ref{sec:the_model_and_}. We find
these issues worth a deeper investigation, which we would like to
devote a future work to.

\appendix

\section{Bethe free energy}
\label{app:bethe}

In this appendix we first give a derivation of the recursion
equation~\eref{eq:rec} as a stationarity condition for the
variational Bethe free energy, and then prove the validity of the
expression~\eref{eq:enlib} for the equilibrium free energy.

Let us consider a Bethe lattice with $c$ coordination number, and
assume that a configuration variable $n_i$ is associated to each
lattice bond~$i$. Let us also assume that the Hamiltonian of the
system is the one given in Eq.~\eref{eq:hamiltonian}, which
includes couplings among the $c$ bonds of each site. Let the
coupling terms be invariant under permutation of the configuration
variables. Expecting a homogeneous thermodynamic state (i.e., that
all local probability distributions are equal), we write the Bethe
free energy per site as
\begin{equation}
  F
  =
  \sum_{\{n_i\}} P_{\{n_i\}}
  \biggr( H_{\{n_i\}} + \ln P_{\{n_i\}} \biggr)
  + \frac{c}{2}
  \sum_n p_n
  \biggr( h_n - \ln p_n \biggr)
  \label{eq:enlibvar}
\end{equation}
where $\{n_i\}$ stands for $n_1,\dots,n_c$, while $p_n$ and
$P_{\{n_i\}}$ denote respectively the single-variable and the
$c$-variable probability distributions. Accordingly, $\sum_n$ and
$\sum_{\{n_i\}}$ denote the sums over possible values of the
configuration variables. Let us notice that, as far as the
entropic part is concerned, the latter term of the Bethe free
energy can be thought of as a correction over the former term,
such that the mean field free energy is recovered, when the joint
probability distribution factorizes. Equilibrium probability
distributions can be determined as minima of the Bethe free
energy, satisfying suitable normalization and compatibility
constraints. By ``compatibility'', we mean that marginalizations
of the joint probability distribution must give the
single-variable distribution, according to the relations
\begin{equation}
  p_{n_i}
  =
  \sum_{\{n_j\}_{j \neq i}} P_{\{n_j\}}
  \ \ \ \ \ \ \ \
  i = 1, \dots, c,
  \label{eq:constraints}
\end{equation}
where the sum runs over possible values of the configuration
variables $n_1,\dots,n_c$, except $n_i$. We thus have a
constrained optimization problem, for which, in the framework of
the Lagrange multiplier method, we can solve analytically
stationarization with respect to probability distributions. Doing
so, the latter can be written as a function of suitable variables
$z$, $Z$, and $w_n$, which correspond to Lagrange multipliers, and
are to be determined in order to satisfy the constraints. We
obtain
\begin{eqnarray}
  p_n
  & = &
  z^{-1} e^{h_n} w_n^2
  ,
  \label{eq:pi} \\
  P_{\{n_i\}}
  & = &
  Z^{-1} e^{-H_{\{n_{i}\}}}
  \prod_{i=1}^c w_{n_i}
  .
  \label{eq:pa}
\end{eqnarray}
Let us notice that $w_n$ plays the role of the (normalized)
partial partition function introduced in the text, whereas the two
constants $z$ and $Z$, associated to the normalization
constraints, are easily determined as a function of~$w_n$.
Moreover, imposing the compatibility
constraints~\eref{eq:constraints}, one obtains the following
recursion equation
\begin{equation}
  w_{n_i} = q^{-1} e^{-h_{n_i}}
  \sum_{\{n_j\}_{j \neq i}} e^{-H_{\{n_j\}}}
  \prod_{\stackrel{\scriptstyle j=1}{j \neq i}}^c
  w_{n_j}
  , \label{eq:m2m}
\end{equation}
where
\begin{equation}
  q = Z/z
  .
  \label{eq:normrec}
\end{equation}
It is possible to show that, because of a slight redundance of the
constraints, one can choose the constant~$q$ at each iteration in
an arbitrary way, for instance by imposing the normalization
condition $\sum_n w_n = 1$, without affecting ``observable''
quantities. Let us also notice that the $c$ compatibility
conditions~\eref{eq:constraints} would require in principle $c$
sets of ``Lagrange multipliers'' $w_n^{(i)}$, for $i=1,\dots,c$.
Nevertheless, one can show that, due to invariance of
$H_{\{n_i\}}$ under permutation, all the sets must be equal to a
single one, which we have just denoted as $w_n$.

Let us now derive the simple free energy formula~\eref{eq:enlib}
presented in the text. Let us plug the expressions~\eref{eq:pi}
and~\eref{eq:pa} for the equilibrium probability distributions
into the logarithmic terms of the variational free
energy~\eref{eq:enlibvar}. By simple algebra, we obtain
\begin{eqnarray}
  F
  & = &
  - \biggr( \sum_{\{n_i\}} P_{\{n_i\}} \biggr) \ln Z
  + \frac{c}{2} \biggr( \sum_n p_n \biggr) \ln z
  \nonumber \\ &&
  - \sum_{i=1}^c \sum_{n_i}
  \biggr( p_{n_i} - \sum_{\{n_j\}_{j \neq i}} P_{\{n_j\}} \biggr)
  \ln w_{n_i}
  .
\end{eqnarray}
Since at equilibrium the normalization and compatibility
constraints are satisfied, the previous expression immediately
simplifies to
\begin{equation}
  F = - \ln Z + \frac{c}{2} \ln z
  .
\end{equation}
Taking into account that there are $c/2$ bonds per site, that
$c=k+1$, and making use of Eq.~\eref{eq:normrec}, we finally
obtain Eq.~\eref{eq:enlib} for the equilibrium free energy per
bond.

\section{Recursion equations}
\label{app:rec}

In this appendix we write an explicit form for the recursion
equations~\eref{eq:rec}, introducing the values of the
couplings~$H_{n_0,\dots,n_k}$, determined according to criteria
explained in Tab.~\ref{tab:energie}. Moving the constant~$q$ and
the single-variable energies~$h_n$ to the left-hand sides, and
remembering that $h_0=0$, we obtain
\begin{eqnarray}
  q w_0
  & = &
  \sum_{\stackrel{\scriptstyle m=0}{m \neq 1,2}}^{k}
  \bin{k}{m} w_2^m w_0^{k-m}
  + \bin{k}{2} e^{\gamma} w_2^2 w_0^{k-2}
\nonumber \\ &&
  + \bin{k}{2} w_1^2 \sum_{m=0}^{k-2} \bin{k-2}{m} w_2^m w_0^{k-2-m}
  ,
  \\
  q e^{h_1} w_1
  & = &
  \bin{k}{1} w_1 \sum_{m=0}^{k-1} \bin{k-1}{m} w_2^m w_0^{k-1-m}
  ,
  \\
  q e^{h_2} w_2
  & = &
  \sum_{m=2}^{k} \bin{k}{m} w_2^m w_0^{k-m}
  + \bin{k}{1} e^{\gamma} w_2 w_0^{k-1}
\nonumber \\ &&
  + \bin{k}{2} w_1^2 \sum_{m=0}^{k-2} \bin{k-2}{m} w_2^m w_0^{k-2-m}
  .
    \label{eq:rec-2prev}
\end{eqnarray}

Let us give a physical explanation of the various terms appearing
on the right-hand sides. They have to take into account all the
allowed configurations of $k$ bonds sharing one site with a given
bond, whose configuration is fixed ($n=0,1,2$ for the three
equations, respectively). In the first equation, the fixed
configuration is $n=0$ (empty bond). In the right-hand side, the
first two terms refer to configurations with $m=0,\dots,k$ bonds
occupied by paired segments, and $k-m$ empty bonds. As explained
in the text, the case $m=1$ is forbidden, and the case $m=2$ is
treated separately, since it has to take into account a stacking
energy contribution. The third term deals with the case of $2$
bonds occupied by unpaired segments, $m=0,\dots,k-2$ by paired
segments, and $k-2-m$ empty bonds. In the second equation, the
fixed bond configuration is $n=1$ (bond occupied by an unpaired
segment). Therefore, in the right-hand side, there is always one
bond occupied by an unpaired segment, together with
$m=0,\dots,k-1$ occupied by paired segments, and $k-1-m$ empty
bonds. In the third equation, the fixed configuration is $n=2$
(bond occupied by paired segments). In the right-hand side, the
first two terms refer to configurations with $m=0,\dots,k$ bonds
occupied by paired segments, and $k-m$ empty bonds. As in the
first equation, the case $m=0$ is forbidden, and the case $m=1$ is
treated separately, because of the stacking energy contribution.
The third term deals with $2$ bonds occupied by unpaired segments,
$m=0,\dots,k-2$ by paired segments, and $k-2-m$ empty bonds.

The above form of the recursion equations can be further
simplified, making use of the binomial expansion. By simple
algebra, we finally obtain
\begin{eqnarray}
  q w_0
  & = &
  (w_0 + w_2)^k
  + \bin{k}{2} w_1^2 (w_0 + w_2)^{k-2}
  \label{eq:rec-0}
  \\ &&
  - \bin{k}{1} w_2 w_0^{k-1}
  + \bin{k}{2} (e^{\gamma}-1) w_2^2 w_0^{k-2}
  ,
  \nonumber \\
  q e^{h_1} w_1
  & = &
  \bin{k}{1} w_1 (w_0 + w_2)^{k-1}
  ,
  \label{eq:rec-1}
  \\
  q e^{h_2} w_2
  & = &
  (w_0 + w_2)^k
  + \bin{k}{2} w_1^2 (w_0 + w_2)^{k-2}
  \nonumber \\ &&
  - w_0^k
  + \bin{k}{1} (e^{\gamma}-1) w_2 w_0^{k-1}
  .
  \label{eq:rec-2}
\end{eqnarray}

\section{Theta point}
\label{app:theta}

Hereafter, we report the derivation of Eq.~\eref{eq:tric}, i.e.,
the analytical expression for the $\Theta$ transition temperature.
Eq.~\eref{eq:crit}, i.e., the second order O-I phase boundary,
comes out as a by-product of this derivation. Let us first define
the ratios $\ra \equiv w_1/w_0$ and $\rb \equiv w_2/w_0$, for
which we can easily derive two recursive equations
from~\eref{eq:rec-0}, \eref{eq:rec-1}, and~\eref{eq:rec-2}
\begin{eqnarray}
  \ra
  & = &
  e^{-h_1}
  \frac
  {
    \bin{k}{1} \ra (1 + \rb)^{k-1}
  }{d}
  ,
  \label{eq:rec-x}
  \\ && \nonumber \\
  \rb
  & = &
  e^{-h_2}
  \frac
  {
    (1 + \rb)^k
    + \bin{k}{2} \ra^2 (1 + \rb)^{k-2}
    - 1 + \bin{k}{1} (e^{\gamma}-1) \rb
  }{d}
  ,
  \label{eq:rec-y}
  \nonumber \\ &&
\end{eqnarray}
where
\begin{equation}
  d
  \equiv
  (1 + \rb)^k
  + \bin{k}{2} \ra^2 (1 + \rb)^{k-2}
  - \bin{k}{1} \rb + \bin{k}{2} (e^{\gamma}-1) \rb^2
  .
  \label{eq:den}
\end{equation}
From~\eref{eq:rec-x} and~\eref{eq:den}, assuming that $\ra \neq
0$, i.e., that we are in the I~phase, we obtain
\begin{equation}
  \bin{k}{2} \ra^2
  =
  \bin{k}{1} e^{\mu} (1+\rb) - (1+\rb)^2
  + \frac{\bin{k}{1} \rb + \bin{k}{2} (e^\gamma-1) \rb^2}{(1+\rb)^{k-2}}
  .
\end{equation}
Moreover, in the $\rb \to 0$ limit, i.e., very close to the
boundary with the O phase, we can write
\begin{equation}
  \bin{k}{2} \ra^2 = \left( k e^{\mu} - 1 \right)
  + \left( k e^{\mu} + k - 2\right) \rb
  + \mathcal{O}(\rb^2)
  .
\end{equation}
Since we have observed from the numerics that such boundary is
second order, we have that $\rb \to 0$ should imply also $\ra \to
0$. We then argue that the zeroth order term on the right-hand
side of the previous equation must vanish. In this way we obtain
Eq.~\eref{eq:crit}. Plugging this equation into the previous one,
we obtain
\begin{equation}
  \bin{k}{2} \ra^2
  =
  (k-1)\rb + \mathcal{O}(\rb^2)
  ,
  \label{eq:xvsy}
\end{equation}
whereas, remembering also Eq.~\eref{eq:h2}, Eq.~\eref{eq:rec-y}
becomes
\begin{equation}
  \rb
  =
  e^\beta \frac{k(e^\gamma+1)-1}{k^2 e^\gamma} \rb + \mathcal{O}(\rb^2)
  .
\end{equation}
Now, in order to allow the possibility that, along the phase
boundary, there can exist some point in which $\rb$ is vanishingly
small but not zero (i.e., a tricritical point), we have to equate
the first order coefficients on the two sides of the previous
equation. By simple algebra, we obtain the $\Theta$ point
condition~\eref{eq:tric}.

\end{document}